\documentclass[a4paper,12pt]{article}

\usepackage{amsmath,amssymb,amsfonts}
\usepackage{graphicx}
\usepackage{color}
\makeatletter
\@addtoreset{equation}{section}
\renewcommand{\theequation}{\thesection.\@arabic\c@equation}
\makeatother
\usepackage{hyperref}
\usepackage{cite}
\usepackage{caption}
\definecolor{red}{rgb}{1,0,0}
\definecolor{green}{rgb}{0,1,0}
\definecolor{blue}{rgb}{0,0,1}
\definecolor{darkblue}{rgb}{0,0,0.5}
\definecolor{lightblue}{rgb}{.5,.5,1}
\definecolor{lightgray}{gray}{.87}
\definecolor{Dark}{gray}{.20}
\definecolor{pink}{rgb}{.95,0.82,0.92}
\definecolor{yellow}{rgb}{1,1,0}
\definecolor{lightyellow}{rgb}{1,1,.5}
\definecolor{purple}{rgb}{0.7,0,0.85}
\definecolor{darkgreen}{rgb}{0,0.5,0}
\definecolor{orange}{rgb}{0.8,0.2,0.2}
\def \be {\begin{equation}}
\def \ee {\end{equation}}
\def \bea {\begin{align}}
\def \eea {\end{align}}
\def \nn {\nonumber}

\def \rr {\raise.35ex\hbox{\small $\prime$}\kern-.17em{\mbox{\large $\imath$}}}

\def \dels {\partial\kern-.5em / \kern.5em}
\def \As {{A\kern-.5em / \kern.5em}}
\def \Ds {D\kern-.7em / \kern.5em}

\def \a {\alpha}

\def \g {\gamma}

\def \d {\delta}

\def \lam {\lambda}

\def \s {\sigma}

\def \om {\omega}

\def \II {I\hspace{-.1em}I\hspace{.1em}}

\newcommand{\detail}[1]{}

\newcommand{\hide}[1]{}

\pdfoutput=1

\begin{document}

\pagestyle{plain}

\begin{titlepage}
\vspace*{-10mm}   
\baselineskip 10pt   
\begin{flushright}   
\begin{tabular}{r} 

\end{tabular}   
\end{flushright}   
\baselineskip 24pt   
\vglue 10mm

\begin{center}

\noindent
\textbf{\LARGE
Final-State Condition And \\
\vskip0.5em
Dissipative Quantum Mechanics
}
\vskip20mm
\baselineskip 20pt

\renewcommand{\thefootnote}{\fnsymbol{footnote}}

{\large
Pei-Ming~Ho
\footnote{pmho@phys.ntu.edu.tw},
}

\renewcommand{\thefootnote}{\arabic{footnote}}

\vskip5mm

{\it
Department of Physics and Center for Theoretical Physics, \\
National Taiwan University, Taipei 106, Taiwan,
R.O.C. 
}

\vskip 25mm
\begin{abstract}

Unitarity demands that
the black-hole final state 
(what remains inside the event horizon
at complete evaporation)
must be unique.
Assuming a UV theory with infinitely many fields,
we propose that the uniqueness of the final state
can be achieved via a mechanism analogous to
the quantum-mechanical description of dissipation.

\end{abstract}
\end{center}

\end{titlepage}

\pagestyle{plain}

\baselineskip 18pt

\setcounter{page}{1}
\setcounter{footnote}{0}
\setcounter{section}{0}


\newpage

\section{Introduction}
\label{introduction}
\label{1}

Recently,
there has been great interest in the information loss paradox of black holes.
A recent development that has attracted a lot of attention is
a new way to calculate the entropy of Hawking radiation
\cite{Penington:2019npb,Almheiri:2019psf}
with results consistent with the Page curve \cite{Page:1993wv,Page:2013dx}.
On the other hand,
the physical mechanism behind the entropy formula remains mysterious,
although it must be some quantum gravity effect
(e.g. microscopic wormholes in the ER = EPR proposal \cite{Maldacena:2013xja}).
As an effort to understand the underlying physics in further detail,
we study in this work a stronger condition of unitarity.

Horowitz and Maldacena proposed \cite{Horowitz:2003he} that
the unitarity of black-hole evaporation can be preserved by
imposing a unique final state as a boundary condition
at the singularity of the gravitational collapse.
It was then challenged \cite{Gottesman:2003up} that
this condition may not be sufficient to ensure unitarity.
Nevertheless,
we now show that a necessary condition of unitarity 
is the uniqueness of the final state of the black hole.

\begin{figure}
\vskip-4em
\center
\includegraphics[scale=0.6,bb=0 -70 200 370]{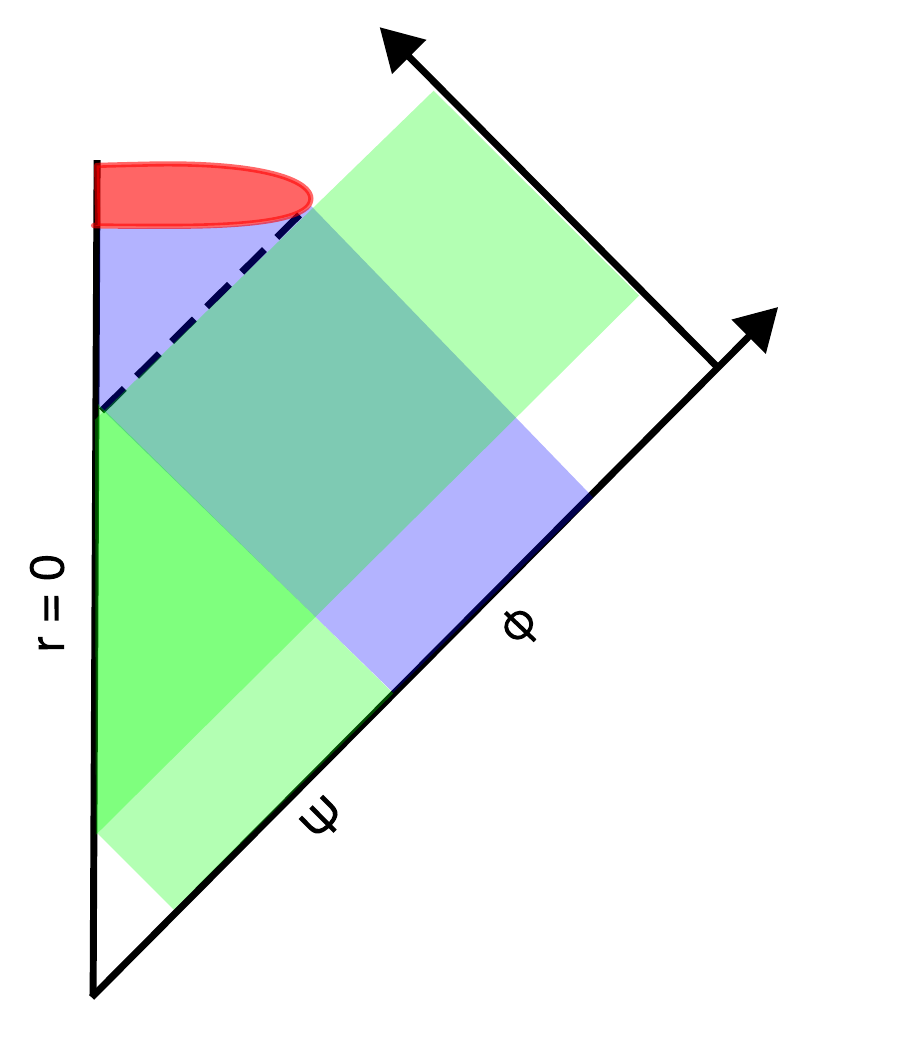}
\vskip-3em
\caption{\small
Dynamical geometry of a black hole. \\
The Hilbert space ${\cal H}_{in}$ describes
ingoing modes (the blue region), 
including the collapsing matter,
that end up at the singularity (red blob).
Outgoing modes in the green region
end up outside the event horizon (dash line)
belong to ${\cal H}_{out}$.
The ingoing and outgoing modes intersect
outside the event horizon.
}
\label{Penrose}
\vskip1em
\end{figure}

The Hilbert space of a black hole is composed of
${\cal H}_{in}$ for the ingoing quantum modes that
eventually end up at the spatial singularity
(a neighborhood of Planckian physics in the UV theory),
and ${\cal H}_{out}$ for the outgoing modes that
eventually come out of the horizon.
(See Fig.\ref{Penrose}.)
\footnote{
The Hilbert space for ingoing (outgoing) modes that
never enter (exit) the horizon is irrelevant.
}
A given pure initial state
$|\Phi_0\rangle = |\phi_0\rangle\otimes|\Psi_0\rangle$
in general evolves to another state 
$|\Phi'\rangle = \sum_n |\phi'_n\rangle\otimes|\Psi'_n\rangle$,
where $|\phi_0\rangle, |\phi'_n\rangle \in {\cal H}_{in}$
and $|\Psi_0\rangle, |\Psi'_n\rangle \in {\cal H}_{out}$.
At the end of the evaporation,
unitarity demands that the Hawking radiation is a pure state by itself,
i.e. the final state is of the form
$|\Phi\rangle = |\phi\rangle\otimes|\Psi\rangle$.

If, for a set of different initial states
$|\Phi^{(n)}_0\rangle = |\phi^{(n)}_0\rangle\otimes|\Psi_0\rangle$,
where $|\Psi_0\rangle$ is the Minkowski vacuum of the infinite past.
the corresponding final states are
$|\Phi^{(n)}\rangle = |\phi^{(n)}\rangle\otimes|\Psi^{(n)}\rangle$,
their superposition gives the time evolution
\begin{align}
|\Psi_0\rangle = 
\left(\sum_n c_n |\phi^{(n)}_0\rangle\right)\otimes|\Psi_0\rangle
\quad \rightarrow \quad
|\Psi\rangle = 
\sum_n c_n |\phi^{(n)}\rangle\otimes|\Psi^{(n)}\rangle.
\end{align}
The entanglement entropy of the black hole would not vanish unless
$|\psi^{(n)}\rangle = |\psi^{(0)}\rangle$ is a {\em unique} state for all $n$
at complete evaporation.
We will refer to this requirement as the {\em final-state condition}.

In a consistent quantum theory of gravity
that respects unitarity and resolves singularities,
the uniqueness of the final state should be achieved
as a result of time evolution,
instead of being imposed as a boundary condition.
What kind of mechanisms can bring
an arbitrary initial state $|\phi_0^{(n)}\rangle$
to the same final state $|\phi^{(0)}\rangle$?
We propose that the answer to this question
is closely related to the familiar phenomena of dissipation.


\section{Dissipative Quantum Mechanics}
\label{2}

A ubiquitous feature of the macroscopic world is dissipation.
A macroscopic oscillator $q$ is better described by the equation
\be
\ddot{q} + \gamma \dot{q} + \om^2 q = 0
\label{dampedosceq}
\ee
with a constant $\gamma > 0$ parametrizing the dampting effect.
Regardless of the initial state,
eventually $q \rightarrow 0$ as $t \rightarrow \infty$.
As the underlying quantum system is unitary,
the information about the initial state cannot be lost.
How is the information preserved in the quantum system?
The answer to this information loss problem is well known
\cite{Callen:1951vq,Weber:1953zz,Senitzky:1958zz,
Senitzky:1960zz,Senitzky:1961zz,Caldeira:1982uj,Leggett:1987zz}.

As a simple example,
we consider the coupling of a simple harmonic oscillator $(q, p)$
to a system of practically infinitely many oscillators $\{q_i, p_i\}$.
The total Hilbert space is ${\cal H}_{total} = {\cal H}_{osc} \otimes {\cal H}_{S}$.
In the Heisenberg picture,
the quantum states are time-independent,
and the operators evolve with time.
It was found that the solutions of $q(t)$ and $p(t)$ 
are given by \cite{Senitzky:1958zz,Senitzky:1960zz,Senitzky:1961zz}
\begin{align}
q(t) = e^{- \g t/2} q_{osc}(t) + \tilde{q}(t),
\qquad
p(t) = e^{- \g t/2} p_{osc}(t) + \tilde{p}(t),
\label{qp-sol}
\end{align}
where
\begin{align}
q_{osc}(t) \equiv q(0)\cos(\om' t) + p(0)\sin(\om' t),
\qquad
p_{osc}(t) \equiv p(0)\cos(\om' t) - q(0)\sin(\om' t),
\label{qp-sol-0}
\end{align}
are the solutions of an undamped oscillator with a shifted frequency $\om'$,
and $\tilde{q}(t)$ and $\tilde{p}(t)$ are operators on ${\cal H}_S$.
The damping parameter $\g$ depends on
the coupling between the oscillator $(q, p)$ and the large quantum system $\{q_i, p_i\}$.
According to eqs.\eqref{qp-sol} and \eqref{qp-sol-0},
$\tilde{q}(0) = \tilde{p}(0) = 0$,
and $q(0)$ and $p(0)$ act only on ${\cal H}_{osc}$.


For a given quantum state
$|\Phi\rangle \equiv |\phi\rangle\otimes|\Psi\rangle \in {\cal H}_{osc} \otimes {\cal H}_{S}$,
where $|\phi\rangle$ is normalized,
the solution \eqref{qp-sol} implies that 
\begin{align}
\langle\Phi| q^m(t)p^n(t) |\Phi\rangle
\rightarrow 
\langle\Psi|\tilde{q}^m(\infty)\tilde{p}^n(\infty)|\Psi\rangle
\qquad (t \rightarrow \infty).
\label{q-ind}
\end{align}
Hence,
the information of the initial state in ${\cal H}_{osc}$ is
no longer accessible through the observables $q(t)$ and $p(t)$
in the limit $t \rightarrow \infty$.

In the Schr\"{o}dinger picture,
the operators are constant,
and the quantum state evolves with time.
Eq.\eqref{q-ind} means that
the initial state $|\phi(0)\rangle$ evolves into
a state that is independent of $|\phi(0)\rangle$,
but depending on $|\Psi(0)\rangle$.

At the same time, 
the unitarity of quantum mechanics ensures that
all the information of the initial state $|\phi(0)\rangle$
is preserved in the final state $|\Psi(\infty)\rangle$
of other oscillation modes.
This point is less emphasized in the literature on dissipative quantum mechanics.
We demonstrate it via a toy model in the appendix.

\section{Unique Final State}
\label{3}

There is no mechanism in low-energy effective theories
to transfer the complete information inside the collapsing matter
into Hawking radiation.
UV physics must become relevant, 
one way or another.
\footnote{
It has been advocated that,
for the sake of unitarity,
there must be ${\cal O}(1)$-correction at the horizon \cite{Mathur:2009hf},
e.g. a firewall \cite{firewall},
which invalidates the effective theory.
For concrete models,
see e.g. the fuzzball proposal \cite{FuzzBall}
and the KMY model \cite{KMY,KMY-1}.
The discussion here is independent of the details of the model.
}
Indeed,
it was recently shown \cite{Ho:2020cbf,Ho:2020cvn} that,
as a manifestation of the trans-Planckian problem \cite{trans-Planckian-1} of Hawking radiation,
low-energy effective theories break down
due to generic higher-derivative interactions.

In this work,
we simply assume that the UV physics is at work
between the ingoing modes (including the collapsing matter)
and the outgoing modes
in a certain neighborhood of the black hole.
Furthermore, 
we assume that there is an infinite spectrum of fields in the UV theory.
Apart from the well-known example of string theory,
the existence of an infinite number of fields appears to be
a salient feature of UV-finite, unitary quantum field theories
that admit a perturbative formulation 
\cite{Krasnikov:1987mz,Moffat:1987yj,Krasnikov:1988pb,
Itzhaki:1994dr,Ho:2004ab,Ho:2009my,Ho:2010jp}.

Note that,
in a perturbative UV theory,
even with the assumption of trans-Planckian scatterings,
it is still unclear how the final-state condition can be satisfied.
In the previous section,
we pointed out a mechanism that brings an oscillator to ``forget''
about its initial state.
We now translate this mechanism to
a UV theory with infinitely many fields.

If we focus on a small neighborhood around a given point in spacetime,
each quantum mode is associated with a pair of conjugate operators
that can be identified with those of an oscillator.
For each ingoing mode of an arbitrary field,
we identify its creation and annihilation operators
with $(q, p)$ in the previous section,
up to a linear transformation.
The operators $\{q_i, p_i\}$ are then identified with
the outgoing modes of all other fields.

The initial state of the ingoing mode is
$|\phi(0)\rangle \in {\cal H}_{osc}$,
on which $(q, p)$ acts.
The initial state $|\Psi(0)\rangle \in {\cal H}_{S}$
should be the Unruh vacuum
(which is the time evolution of the Minkowski vacuum of the infinite past).

The discussion in the previous section implies that
the initial state $|\phi(0)\rangle\otimes|\Psi(0)\rangle$ of the black hole
evolves towards a state $|\phi(\infty)\rangle\otimes|\Psi(\infty)\rangle$
with $|\phi(\infty)\rangle$ independent of $|\phi(0)\rangle$.
Since $|\Psi(0)\rangle$ is always chosen to be the Unruh vacuum,
the final state $|\phi(\infty)\rangle$ is unique.
The final-state problem would be solved
(see the appendix for a toy model)
except that the spacetime geometry is cut off
by the space-like singularity at the origin.
As we are not sure how the singularity would be resolved in the UV theory,
we restrict ourselves to $t \leq T$,
with $t = T$ begin a time slice just before the singularity emerges.

The characteristic time scale $\g^{-1}$ of this dissipative mechanism is
typically of the same order of magnitude as
the characteristic length scale of the UV theory (quantum gravity)
which is the Planck length $\ell_p$.
As long as the time span of the region where UV physics is at work
is much longer than $\ell_p$,
in the Heisenberg picture,
eq.\eqref{qp-sol} implies that
\begin{align}
q(T) \simeq \tilde{q}(T),
\qquad
p(T) \simeq \tilde{p}(T).
\label{qp-T}
\end{align}

The same argument applies to all ingoing modes.
We can talk about all the ingoing modes at the same time.
The tensor product of ${\cal H}_{osc}$ for all ingoing modes is ${\cal H}_{in}$,
which is defined in Sec.\ref{1},
and ${\cal H}_S$ can be identified with ${\cal H}_{out}$.
Let $P_{in}$ and $P_{out}$ denote the projections
onto the Hilbert spaces ${\cal H}_{in}$ and ${\cal H}_{out}$, respectively.
Eq.\eqref{qp-T} means that,
in the Schr\"{o}dinger picture,
the state $P_{in}|\Phi(T)\rangle$ of the ingoing modes
is approximately independent of $|\phi(0)\rangle$.
As it can only depend on $|\Psi(0)\rangle$,
which is unique,
the final state $P_{in}|\Phi(T)\rangle$ is approximately unique.
\footnote{
The information of the initial state $|\phi(0)\rangle$
is preserved in the state $P_{S}|\Psi(T)\rangle$
of the outgoing modes.
}

Notice that,
as we have defined the final state at a time $T$ before the singularity emerges,
the singularity at $r = 0$ is not directly related to the final-state problem.
Furthermore,
all we need is approximate uniqueness of the final state,
as the singularity should be resolved in the UV theory
as a Planckian neighborhood which may still carry some information.



Apart from whether we need it to solve the final-state problem,
dissipation is a robust feature of any system
with infinite degrees of freedom.
The evidence is the simple fact that dissipation (e.g. friction) appears everywhere.
It is sufficient to have a UV region much larger than the Planck length,
where UV-theory interactions have to be treated as a continuous effect,
in contrast with brief interactions in scattering processes.

\section{Comments}

We answer some of the natural questions here.

\noindent
{\em
How is this work different from
the proposal of Ref.\cite{Horowitz:2003he}?
}

Our proposal is closely related to,
but quite different from,
the proposal of the final-state boundary condition \cite{Horowitz:2003he}.
We propose here that the final state of the collapsing matter
is unique as the result of a robust feature of UV theories
with infinitely many fields
(while a spectrum of infinitely many fields is argued to be 
a common feature of perturbative formulation of UV theories).
The information is transferred through causal, local interactions.
But it is necessary to assume that
low-energy effective theory breaks down
and UV physics is important somewhere other than the singularity.

On the other hand,
in the proposal of Horowitz and Maldacena,
the final-state boundary condition is imposed by hand
at the space-like singularity of a Schwarzschild black hole,
presumably as an effective description of certain unknown UV physics.
As opposed to our proposal,
the infalling negative-energy Hawking radiation behind the horizon
plays the crucial role of a medium for the post-selected teleportation
of quantum information in their theory,
and everywhere other than the singularity at the origin can be smooth and uneventful.
However, in this scenario,
causality is at risk \cite{Gottesman:2003up,Lloyd:2013bza},
and it is not clear whether
it can be justified by certain UV physics.

\noindent
{\em
What do we really mean by ${\cal H}_{in}$?
}

In our statement of the final-state condition,
we have implicitly assumed that the Hilbert space
can be decomposed into two parts:
${\cal H}_{in}$ is the Hilbert space for everything
that ends up inside the event horizon,
and ${\cal H}_{out}$ that for everything 
that ends up outside the event horizon.
Strictly speaking, 
the existence of the event horizon is uncertain,
as the singularity at the center of the black hole
is expected to be resolved in the UV theory.
Nevertheless,
when the black hole evaporates to $1/n$ of its initial mass,
the entanglement entropy of the state in ${\cal H}_{in}$
should be $\sim \mathcal{O}(2/n)$ for a large $n$
according to the Page curve \cite{Page:1993wv}.
This means that the part of the quantum state in ${\cal H}_{in}$
deviates from the unique state only by a tiny fraction.

\noindent
{\em
What about charged black holes?
}

In general,
the final state depends on conserved charges of local gauge symmetries.
The black hole approaches to a final state of extremal black hole
of the charge specified by the initial state.
The superposition of initial states of different charges can be ignored
due to superselection rules.

\noindent
{\em
Do we really need infinitely many fields?
}

In the analogue between quantum mechanics (QM) and quantum field theories (QFT),
a single quantum field is equivalent to
infinitely many quantum mechanical degrees of freedom.
Does this mean that a single field can do the job of
the infinitely many particles in dissipative quantum mechanics?
In the perturbative formulation of QFT,
a particle in an asymptotic state moves freely until
it is spatially close to another particle.
There is no continuous interaction with other degrees of freedom
as depicted in the dissipation process
unless it passes through a high density of particles.
This is why we do not see the dissipative effect at low energies.

A single field would be sufficient if the region
with a high density of particles is sufficiently large
to complete the dissipation process.
On the other hand,
if the high-density region is very small (e.g. Planck scale),
we do need many fields to enhance the dissipative effect.
\hide
{\color{blue}
The paper should be rewritten so that this point is made clear
from the beginning.
Maybe moving this part above?
}

\noindent
{\em
Does the same mechanism apply to different models of black holes?
}

For the spacetime geometry of the conventional model of black holes,
the evaporation time scale $\mathcal{O}(a^3/\ell_p^2)$ for a distant observer
is causally matched with the proper time scale $\mathcal{O}(\ell_p)$ 
from the viewpoint of a freely falling observer comoving
with the collapsing matter \cite{ShortDistance,Ho:2019pjr}.
(Here $a$ is the Schwarzschild radius and $\ell_p$ the Planck length.)
For the KMY model,
the collapsing matter's proper time scale of evaporation is $\mathcal{O}(a)$.
It is still a very short time scale to burn out all the collapsing matter
unless we have interactions at the Planck scale.
The mechanism outlined above provides such a violent mechanism
applicable to both models.

In the conventional model of black holes,
the collapsing matter passes through the horizon without ``dramma''.
It is still a mystery how a UV theory is relevant
in view of the decoupling principle.
This is a key problem in the information loss paradox
that this paper has not provided any clue.
Instead, 
we comment on the KMY model \cite{KMY,KMY-1}.
In the KMY model,
UV physics is needed at the outer layers of the collapsing matter,
due to a Planckian pressure associated with outgoing radiation.
Matter evaporates layer by layer from the outside.
But how matter is completely turned into radiation
cannot be fully explained in the low-energy effective theory.
Our proposal provides a UV mechanism behind this process,
and tells us that it only takes $\Delta t \sim$ several $\ell_p$
to incinerate each layer.

\noindent
{\em
Conclusion
}

In the above,
we have proposed that the same mathematics of dissipative quantum mechanics
provides a mechanism for UV theories with infinitely many fields
to satisfy the final-state condition
and to transfer the information of the collapsing matter to outgoing radiation.
The mechanism behind this proposal is robust,
applicable to almost all models of black-hole evaporation
as long as the UV theory of quantum gravity
consists of an infinite number of fields.

To conclude,
we have shown that
the quantum mechanical origin of
the classical phenomenon of dissipation
helps resolve the final-state problem of black holes.
It also explains how the information
of the collapsing matter is transferred 
at the same time into outgoing UV modes.

\section*{Acknowledgement}

We thank Hsin-Chia Cheng, Hsien-chung Kao,
Hikaru Kawai, Samir Mathur, Yoshinori Matsuo, and Yuki Yokokura
for valuable discussions.
This work is supported in part by the Ministry of Science and Technology, R.O.C. 
and by National Taiwan University.

\appendix

\renewcommand{\theequation}{A.\arabic{equation}}

\section*{Appendix: A Toy Model}

We construct a toy model of dissipative quantum mechanics
suitable for its application to black holes.
The same qualitative features are argued in the literature
to persist in more general models
\cite{Callen:1951vq,Weber:1953zz,Senitzky:1958zz,
Senitzky:1960zz,Senitzky:1961zz,Caldeira:1982uj,Leggett:1987zz}.

Consider the coupling of a harmonic oscillator $(q, p)$ 
to a system of infinitely many oscillators $\{q_n, p_n\}_{n\in\mathbb{Z}}$
with the following Lagrangian
\begin{align}
L = \frac{1}{2}m\dot{q}^2(t) - \frac{1}{2} m\om^2 q^2(t)
+ \lambda(t) q(t) \dot{q}_0(t)
+ \sum_{n\in\mathbb{Z}} \left(
\frac{1}{2} \dot{q}_n^2(t) - \frac{1}{2} M^2 q_n^2(t)
+ \tilde{\lam} q_n(t)q_{n+1}(t)
\right).
\nn
\end{align}
\detail{
The Lagrangian can be rewritten as
\begin{align}
L = \frac{1}{2}m\dot{q}^2(t) - \frac{1}{2} m\om^2 q^2(t)
+ \lambda(t) q(t) \dot{\Psi}_0(t)
+ \sum_{n\in\mathbb{Z}} \left(
\frac{1}{2} \dot{q}_n^2(t) - \frac{1}{2} \tilde{\lam} (q_{n+1}(t) - q_n(t))^2
- \frac{1}{2}\tilde{M}^2 q_n^2(t)
\right),
\nn
\end{align}
where $\tilde{M}^2 \equiv M^2 - 2\tilde{\lam}$.
}
Without loss of generality,
we set $m = 1$.
For simplicity in the calculation below,
we choose $M^2 = 2\tilde{\lam} > 0$.
In the continuum limit
\begin{align}
n \rightarrow \tilde{\lam}^{1/2}\s,
\qquad
q_n \rightarrow \tilde{\lam}^{-1/4} \psi(t, \s),
\end{align}
the Lagrangian becomes
\begin{align}
L = \frac{1}{2}\dot{q}^2(t) - \frac{1}{2} \om^2 q^2(t)
+ \lambda(t) q(t) \dot{\psi}(t,0)
+ \int_{-\infty}^{\infty} d\s \, \left(
\frac{1}{2} \dot{\psi}^2(t,\s) - \frac{1}{2} {\psi'}^2(t,\s)
\right).
\end{align}
We will now analyze the dissipation mechanism of this model.

The Euler-Lagrange equations are 
\begin{align}
\ddot{q}(t) + \om^2 q(t) - \lam(t) \dot{\psi}(t, 0) &= 0,
\label{eq-q}
\\
\ddot{\psi}(t, \s) - {\psi''}(t, \s)
+ \frac{d}{dt}\left(\lam(t) q(t)\right) \delta(\s) &=0.
\label{eq-Psi}
\end{align}
The general solution of eq.\eqref{eq-Psi} is
\begin{align}
\psi(t, \s) 
= \psi_h(t, \s) - \frac{1}{2} \lam(t - |\s|)q(t - |\s|),
\label{psi-sol-0}
\end{align}
where the homogeneous solution
$\psi_h(t, \s) \equiv \psi_+(t+\s) + \psi_-(t-\s)$
for arbitrary functions $\psi_{\pm}$.
Plugging eq.\eqref{psi-sol-0} into eq.\eqref{eq-q},
we find
\begin{align}
\ddot{q} + \frac{1}{2} \lam^2(t)\dot{q}(t) + \hat{\om}^2(t) q(t) 
= \lam(t)\dot{\psi}_h(t, \s),
\label{q-eom}
\end{align}
where $\hat{\om}^2 \equiv \om^2 + \frac{\lam(t)\dot{\lam}(t)}{2}$.
This is the equation of a damped oscillator.
Its general solution is
\begin{align}
q(t) &= q_h(t) + \tilde{q}(t).
\label{q=qh+tq}
\end{align}
The homogeneous solution $q_h(t)$
describes an oscillation with an exponentially decaying amplitude.
The special solution is
\begin{align}
\tilde{q}(t) &\equiv \int_{-\infty}^{\infty} dt' \, K(t, t') \lam(t') \dot{\psi}_h(t', 0),
\label{tildeq}
\end{align}
where the retarded Green's function is
\begin{align}
K(t, t') \equiv \Theta(t-t') \, e^{-\int_{t'}^t\frac{\lam^2(t'')}{4}dt''} \, \frac{\sin(\om (t-t'))}{\om},
\end{align}
where $\Theta$ is the step function.
\detail{
It satisfies
\begin{align}
\left[\frac{d^2}{dt^2} + \frac{1}{2}\lam^2(t)\frac{d}{dt} + \left(\om^2 + \frac{1}{2}\lam(t)\dot{\lam}(t)\right)\right]
K(t, t') = \d(t - t')
\end{align}
}
The derivation of the solution for $p(t)$ is straightforward.

\detail{
Let $\a_+, \a_-$ denote the roots of $\a^2 + \frac{\lam_0^2}{2}\a + \om^2 = 0$,
i.e.
\begin{align}
\a_{\pm} \equiv - \frac{\lam^2(t)}{4} \pm i \om.
\end{align}
The general homogeneous solution $q_h(t)$ is
\begin{align}
q_h(t) &= A_+ e^{\int_{0}^t \a_+(t') dt'} + A_- e^{\int_{0}^t \a_-(t') dt'}.
\end{align}
Clearly,
\begin{align}
q(t = 0) = q_h(0) = A_+ + A_-,
\qquad
q(t \rightarrow \infty) = \tilde{q}(t).
\end{align}
}

When we apply this toy model to the evaporating black hole,
we identify $(q, p)$ with any infalling propagating quantum mode,
and identify $\{q_n, p_n\}$ (or $\{\psi, \dot{\psi}\}$)
with the outgoing propagating modes of all (infinitely many) fields.
Around a given time during the evaporation of the black hole,
the UV theory is needed only in a certain neighborhood of spacetime.
The coupling $\lam(t)$ in the UV theory is thus turned on only when
the ingoing and outgoing fields intersect in this neighborhood.
Without loss of generality,
we turn on the coupling $\lam(t)$ at $t = 0$.
That is, 
\begin{align}
\lam(t) = \lam_0 \Theta(t).
\end{align}
The solutions of $q$ and $\psi$ are now given by
\begin{align}
q(t) &= q_h(t) 
+ \lam_0 \int_0^t dt' \, e^{-\frac{\lam_0^2(t - t')}{4}} \frac{\sin(\om(t - t'))}{\om}\dot{\psi}_h(t', 0),
\label{q-sol}
\\
\psi(t, \s) &= \psi_h(t, \s) - \frac{1}{2} \lam_0 \Theta(t - |\s|) q(t - |\s|),
\label{psi-sol}
\end{align}
where $q_h(t)$ is given by eq.\eqref{qp-sol} with
\begin{align}
\g = \frac{\lam_0^2}{2},
\qquad 
\om' = \sqrt{\om^2 - \frac{\lam_0^4}{16}}.
\end{align}

In the Heisenberg picture,
the state
$|\phi\rangle\otimes|\Psi\rangle \in {\cal H}_{osc}\otimes{\cal H}_S$
of the quantum system
is constant and the operators evolve with time.
As $q = q_h$ and $\psi = \psi_h$ at the initial time $t = 0$,
the Hilbert spaces ${\cal H}_{osc}$ and ${\cal H}_S$
are representations of $(q_h, \dot{q}_h)$ and $(\psi_h, \dot{\psi}_h)$,
respectively.
At large $t$ ($t \gg 1/\lam_0^2$),
$q(t)$ becomes independent of $q_h$,
hence it can no longer extract information
of the state $|\phi\rangle \in {\cal H}_{osc}$.
Instead,
$q(t)$ becomes dominated by the contribution of $\psi_h$ at large $t$.
In the Schr\"{o}dinger picture,
this means that
the state $|\phi(t)\rangle$ of the collapsing matter becomes
independent of its own initial state $|\phi(0)\rangle$ at large $t$,
but it would be determined by the initial state $|\Psi(0)\rangle \in {\cal H}_{S}$
of the outgoing UV fields.
As the initial state $|\Psi(0)\rangle$ is fixed (the Unruh vacuum),
the final state $|\phi(\infty)\rangle$ is unique.

Another conclusion we draw from the general solution
\eqref{q-sol}, \eqref{psi-sol} is the following.
In the Heisenberg picture,
although the state $|\phi\rangle$ becomes inaccessible
to the operators $(q(t), p(t))$ at large $t$,
it is still accessible at any time $t$
through the operators $\psi(t, \s = \pm t)$ and $\dot{\psi}(t, \s = \pm t)$,
which include $q_h$ and $\dot{q}_h$ without
the exponential suppression factor $e^{- \lam_0^2 t/4}$.
In the Schr\"{o}dinger picture,
this means that
the information of the initial state $|\phi(0)\rangle$ 
is preserved in the state $|\Psi(t)\rangle \in {\cal H}_S$ at large $t$.


\end{document}